\documentclass[a4paper,fleqn,usenatbib]{mnras}

\usepackage[T1]{fontenc}
\usepackage{ae,aecompl}

\usepackage{graphicx}	
\usepackage{amsmath}	
\usepackage{amssymb}	
\usepackage{array}

\title[Deuterated formaldehyde in HH212]{Deuterated Formaldehyde in the low mass
protostar HH212}

\author[Dipen Sahu]{
Dipen Sahu,$^{1,4}$\thanks{E-mail: sahudipen@prl.res.in}
Y.C Minh,$^{2}$
Chin-Fei Lee,$^{3}$
Sheng-Yuan Liu,$^{3}$
Ankan Das,$^{4}$ 
\newauthor S.K Chakrabarti,$^{4,5}$ Bhala Sivaraman.$^{1}$
\\
$^{1}$Physical Research laboratory, Navrangpura, Ahmedabad, Gujarat 380009, India. \\
$^{2}$Korea Astronomy and Space Science Institute, 776 Daedeok-daero, Yuseong, Daejeon 34055, Republic of Korea.\\
$^{3}$Academia Sinica Institute of Astronomy and Astrophysics, P.O. Box 23-141, Taipei 106, Taiwan.\\
$^{4}$Indian Centre for Space Physics, 43 Chalantika, Garia St. Road, Kolkata-700084, India.\\
$^{5}$S.N. Bose National Center for Basic Sciences, JD-Block, Salt Lake, Kolkata 700098, India.}

\date{Accepted XXX. Received YYY; in original form ZZZ}

\pubyear{2017}

\begin{document}
\label{firstpage}
\pagerange{\pageref{firstpage}--\pageref{lastpage}}
\maketitle

\begin{abstract}
HH212, a nearby (400 pc) object in Orion, is a 
Class 0 protostellar system with a Keplerian disk and collimated bipolar SiO jets.
 Deuterated water, HDO  and a deuterated complex molecule, methanol (CH$_2$DOH)
have been reported in the source. Here,
we report the HDCO (deuterated formaldehyde) line observation from ALMA data to probe
the inner region of HH212. We compare HDCO line with other molecular lines to understand
the possible chemistry and physics of the source. The distribution of HDCO emission suggests 
it may be associated with the base of
the outflow. The emission also shows a rotation but it is not associated with the
Keplerian rotation of disk or the rotating infalling envelope, rather it is associated with the outflow
as previously seen in C$^{34}$S. From the possible deuterium fractionation,
we speculate that the gas phase formation of deuterated formaldehyde is active in the central hot
region of the low-mass protostar system, HH212.
\end{abstract}




\section{Introduction}


Stars form mainly due to gravitational contraction of a molecular cloud.
Molecular cloud is like a vast laboratory for space chemistry. From the verge 
of collapse to star formation, a molecular cloud goes through different stages. 
Chemical changes during this evolution acts as 
a hint to understand physical and chemical processes associated with the source. 
For example, SiO is a good shock tracer, high-velocity CO line wings are a good tracer 
of outflows; CH$_3$OH can trace other kinds of environments, including outflows
\citep{taquet2015} and even cold gas in pre-stellar cores \citep{vastel2014}, C$^{17}$O helps to 
know disk properties in some cases, etc. After the collapse of a molecular cloud, 
a young stellar object forms at the central region. This central region is embedded in a thick rotating 
envelope which further forms a disk like region. Some of the matter is often ejected in the form of bipolar outflow
carrying significant amount of angular momentum.
\citet{lada87} classified an evolutionary sequence of young stellar objects depending on the spectral energy 
distribution, e.g., class I, class II, class III. Younger protostars are often called class 0 object \citep{andre93}.
HH212 is thought to be a Class 0 object of low-mass proto stellar system, located in Orion at 400pc \citep{koun2017}.

The inner region of the low-mass star-forming cores may be enriched with complex molecules, e.g.,
methyl formate (HCOOCH$_3$), ethyl cyanide ($\mathrm{C_2H_5CN}$), dimethyl ether 
($\mathrm{CH_3OCH_3}$), methyl alcohol ($\mathrm{CH_3OH}$) and formaldehyde (H$_2$CO) \citep{biss2008, bott2004, cazaux2003}.
To differentiate between the `hot core' region that are present in high mass star 
forming region, a term `hot corino' \citep{ceccar2007} is often used in the context of a low mass 
star forming region. Molecules can form both in the gas phase and on the grain surface --
during the evolution of molecular cloud in gas phase, 
molecules can deplete on grain surface and start new chemistry there. 
Grain surfaces help formation of complex molecules by hydrogenation; 
there are also routes by which complex molecules can
also form in the gas phase too. Complex molecules like HCOOCH$_3$,
CH$_3$OCH$_3$ \citep{balu2015}, NH$_2$CHO \citep{barone2015}
can effectively form in gas-phase following sublimation of key
simpler precursors, e.g., CH$_3$OH, NH$_2$, HDCO from grain mantles.
Complex molecules which form on grain mantles can populate the gas-phase 
by desorption from grain mantels. Due to thermal
desorption in high-temperature region e.g., hot core, hot corinos,
complex molecules or their precursors come out of the grain surfaces in gas phase.
For this reason, high deuteration of complex species in grain surfaces
may be reflected to gas phase (e.g., \citealp{cc2014, das2015a, das2015b}). Deuterated water
 has been reported by \citet{code2016} in the hot-corino of HH212. Very recently 
\citet{lee2017a} reported D$_2$CO and singly deuterated methanol (CH$_2$DOH),
first deuterated complex molecule to be observed in central
region of HH212.

Due to limited spatial resolution of observational facilities in the past, only a few hot-corinos have
been reported till now, on $<$100 AU scale, e.g., IRAS 16293-2422, NGC 1333 IRAS2A, IRAS4A etc.
(e.g., \citealp{imai16,jorgen2012} and references therein).
Recently, \citet{code2016} has suggested a `hot-corino' region in HH212
system. The HH212 region is an ideal system to investigate different processes (e.g.,
infall, rotation, hot corino, bipolar outflow) related to the formation of a low mass protostar. 
The HH212 source has been observed in the past using Sub-millimeter array 
(SMA)\citep{lee2006}, the IRAM Plateau de Bure (PdB) interferometer \citep{code2007} and Atacama Large Millimeter Array (ALMA)
\citep{lee2014, code2014}. ALMA, due to its high spatial resolution and sensitivity has revealed much more 
detail than any other previous observations. The HH212 system has bipolar jets as observed by SiO, SO, and SO$_2$ 
emission lines \citep{code2014,podio2015}. It has a central hot-corino surrounded by a flattened, 
infalling and rotating envelope as observed by 
C$^{17}$O and HCO$^+$ \citep{lee2006,lee2014}. Also, from HCO$^+$ and C$^{17}$O emission line 
observations, \citet{lee2014}, \citet{code2014} suggested a compact disk ($\sim 90$AU) rotating around the 
source of $\cong 0.2-0.3$M$\odot$.  Later observation \citep{lee2017b} resolved the disk and
suggested the disk size to be 40 AU.

In this paper, we use ALMA archival dataset 2011.0.00647.S 
and report emission of deuterated formaldehyde line from central hot-corino region. We compare the HDCO emission 
with methanol (CH$_3$OH), C$^{34}$S and C$^{17}$O and try to explain chemistry around the central region of the source.
The line search is performed using splatalouge (www.splatalouge.net) and molecular data has been taken from
the CDMS \citep{muller01,muller05} and JPL molecular database \citep{pic1998}.

\section {Observations}
The HH212 protostar system was observed with ALMA (Band -7) using 24 12m antennas on 2012 December 1 
(Early Science Cycle 0 phase
, \citealt{code2014}). In this observation, the shortest and the longest baselines were respectively 20m and 360m. We report 
HDCO (Table~\ref{tab:lines}) and include the CH$_3$OH, C$^{34}$S and C$^{17}$O lines 
to compare with deuterated formaldehyde emission. The datacubes have a spectral resolution 
488 KHz ($\sim 0.43$ km s$^{-1}$), a typical beam FWHM of $0.''65 \times 0.''47$ at position angle (PA) $\sim 49^{\circ}$.
The observed spectral windows were 333.7-337.4 GHz and 345.6-349.3 GHz, the typical rms level was 3-4 mJy beam$^{-1}$
in 0.43 km s$^{-1}$ channels.
The data were calibrated with the CASA package, with
quasars J0538-440 and J0607-085 as the passband calibrators, quasar
J0607-085 as the gain calibrator, and Callisto and Ganymede as the
flux calibrators. We generated spectral cubes by subtracting the continuum emission in visibility data.
We used Briggs
weighting with robustness parameter 0.5 for CLEANing the image.
Positions are given with respect to peak continuum of the MM1 protostar located at 
$\alpha(J2000)= 05^h43^m51^s.41,
 \delta(J2000)=-01^{\circ}02'53''.17$ \citep{lee2014}.

\section{Results and Discussions}



\begin{figure}

\includegraphics[width=\columnwidth]{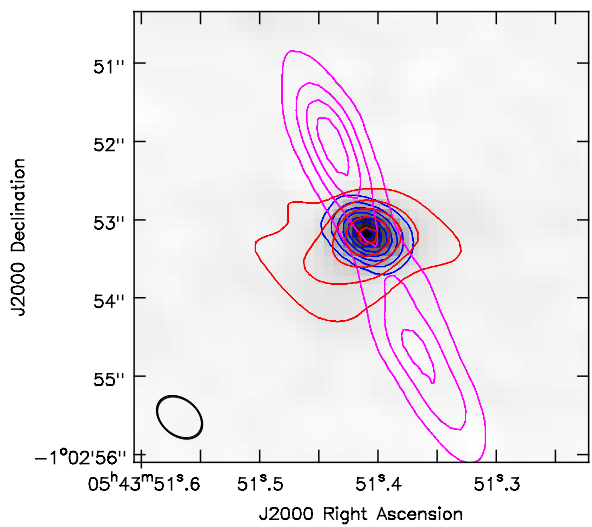}

\caption{HH212 system, observed by ALMA -Band 7 \citep{code2014}. Here, integrated emissions (moment 0 map) of three
     lines, SiO, HDCO and C$^{17}$O are overlaid on source continuum (gray scale). 
     Magenta contours show the SiO(8-7) bipolar jet; 
     first contour and steps are 10$\sigma$
     ($\sigma\sim$ 5 mJy/beam km/s).
     Blue contour is of HDCO; contours are from 7$\sigma$ and steps are 8$\sigma$(15 mJy/beam km/s).
     Red contour is of C$^{17}$O, contours are from 5$\sigma$ in steps of 5$\sigma$ (1.5 mJy/beam km/s)}
     
\label{fig:hdco_overlap}
\end{figure}

ALMA band 7 is found to be rich in numerous spectral transitions in a region 
towards MM1 protostar. Here, we discuss 
three lines, C$^{17}$O, deuterated formaldehyde line (HDCO 5(1,4)-4(1,3)) and methanol (CH$_3$OH(v0 7(1,7)- 6(1,6)).
The details of the lines are given in Table~\ref{tab:lines}. The C$^{17}$O (3-2)  line transition was used by \citet{code2014}
to describe the rotating envelope around the central region. Here, we consider C$^{17}$O (3-2) 
emission as optically thin and have calculated 
its column density around different regions of the source. We compare C$^{17}$O column density with HDCO and CH$_3$OH
column density to study the chemistry around the MM1 protostar.
From Figure~\ref{fig:hdco_overlap}, HDCO emission is seen in central region of the protostar system and 
partially resolved (synthesized beam size $0.66'' \times 0.47'', {\rm PA}\quad 49.6^{\circ}$ and image size $0.70''
\times 0.49'', {\rm PA}\quad 57.2^{\circ}$).
For the first time, we are reporting deuterated formaldehyde emission around the MM1 protostar position.
Recent work by \citet{leurini2016} 
has described the kinematics of methanol in HH212. Moreover, from a recent observation (ALMA, 2015)
\citet{lee2017a} found that methanol is from a rotating disk environment. Here, we compare formaldehyde emission 
with the most intense methanol line (E$_u$=79 K). Both the line profiles of formaldehyde and methanol
have peaks at $\sim$ 1.9 km s$^{-1}$, i.e., close to the systematic velocity $\sim$1.7 km s$^{-1}$ \citep{lee2014}.
Though \citet{code2014} suggested a systematic velocity of $\sim$ 1.3 km s$^{-1}$,
here we consider the systemic velocity to be $\sim$ 1.7 km s$^{-1}$. 

\begin{table}
	\centering
	\caption{List of unblended transitions detected towards HH212-MM1 and line properties.}
	\label{tab:lines}
	\begin{tabular}{lccc c} 
		\hline
		Species & Transition& Frequency& $E_u$& S$\mu^2$\\
		        &        & (GHz)       &(K)    & D$^2$  \\
		\hline
		CH$_3$OH &v0 7(1,7)- 6(1,6) & 335.58202 &78.97& 5.55\\ 
		HDCO & 5(1,4)-4(1,3)& 335.09678 &56.25&26.05422\\
		C$^{17}$O & J=3-2 & 337.06112&  32.35  & 0.01411\\
		C$^{34}$S & J=7-6 & 337.396459 &   50.23 & 25.57\\
		\hline
	\end{tabular}
\end{table}

\subsection{HDCO emission}
The HDCO emission is almost symmetric around the systematic velocity ($\sim 1.6-2.0$ km s$^{-1}$), so it can be assumed that 
it originates near the central hot region. In this region, the dust temperature is 
high enough, and molecules on the grain surface (i.e., on the dust) easily desorb to gas phase. 
Temperature of hot-corinos may 
vary from few 10's K to few 100's K \citep{parise2009, code2016}. Here, we have observed only one HDCO transition, 
so the temperature can not be derived directly. \citet{code2016} used the same dataset and listed five acetaldehyde 
(CH$_3$CHO) transitions; from optically thin LTE analysis of CH$_3$CHO they suggested a temperature of $87\pm47$K.
\citet{leurini2016} reported a rotational temperature 295 K for methanol emissions, while 
\citet{lee2017a} finds an excitation temperature of 165$\pm$85 K for deuterated methanol transitions.
Considering all these results and roughly the temperature variation along the radius \citep{lee2014}
we have considered five excitation temperatures (20, 40, 90, 160, 300 K) for our calculations.
We consider some low excitation temperatures e.g., 20 \&
40 K for regions away from hot-corino and high-temperatures for a hot region.
 Here we assume the HDCO emission region to be optically thin and local thermodynamic equilibrium (LTE) condition
is satisfied.
The details of column density calculation  and the variation of it, is described in Section 3.3.

In Figure~\ref{fig:hdco_overlap}, we see that HDCO emission is concentrated in a circular
region around the central protostar position, i.e., the peak of continuum emission. The 
emission is elongated along the jet direction. 
As the emission is only partially resolved, we can not infer conclusively whether the elongation 
is real or because of the effect of synthesized beam size. Figure 2 shows that near the systematic
velocity, the HDCO emission is most extended; though the extended emission feature is weak ($\sim 3\sigma$), 
it is absent in higher velocity channels from the systematic velocity. 
This extended emission feature is similar to the `X' shaped outflow cavity as traces by C$^{34}$S emission.
To draw a further conclusion, in Figure~\ref{fig:channel_map} we plot three velocity channels near
the systematic velocity
for HDCO emission and compare it with C$^{34}$S channel maps. The C$^{34}$S emission is 
tracing a dense gas component. `X' shaped outflow is closely related to the bipolar jet or outflow
near the systematic velocity \citep{code2014}. From 
Figure ~\ref{fig:channel_map}, it can be seen that though the emissions of HDCO away from the central source
is weak, it is significantly similar to the C$^{34}$S emission near the systematic velocity. 
To find whether HDCO emission shows any rotation, we plot centroid emission positions of various
velocity channels of C$^{34}$S and HDCO line in Figure ~\ref{fig:uv1}.  \citet{code2014} finds an
evident rotation around jet for C$^{34}$S  emission in the southern lobe. From 
Figure ~\ref{fig:uv1} we can see that the there is definite signature of rotation for 
C$^{34}$S as the blue shifted and red-shifted emission are situated away from the 
central peak position and the jet axis. 
At low velocities ($<1.0$), it is clearly showing rotation (in southern lobe) as emission
centroids of blue-shifted and red-shifted emission situated roughly symmetrically away from
the jet-axis and below the disk plane. At a high velocity, the rotation feature is not 
clear but becomes more collimated      
towards the jet axis. The HDCO emission centroids also show a  signature of rotation which is
clear from Figure~\ref{fig:uv1}. Looking to the southern portion 
of emission there seems to be some similarity with 
C$^{34}$S features. As the intensity of emission away from the source is very faint for HDCO 
and the shift of different velocity channels is less than the beam size, this inference 
may not be conclusive. We can speculate that as the emission centroid (red-shifted southern
portion in Fig. 4) of HDCO is away from the disk plane and shifts along the jet axis. So, it 
is not associated with disk rotation; at smaller scales, it may be associated with disk wind 
or small-scale outflow or cavity rotation but with current spatial resolution of the ALMA data we cannot confirm it.
\citet{leurini2016} finds that methanol (CH$_3$OH) could trace the base of the low-velocity of the small 
scale outflow and another higher resolution observation (0.04$''$, \citealp{lee2017a}) suggests that
it is from a warm environment near the disk surface. Hence, to discuss the origin HDCO of emission, we 
compare HDCO emission with CH$_3$OH and C$^{17}$O emission in the next section. 

\begin{figure}
	\includegraphics[width=\columnwidth]{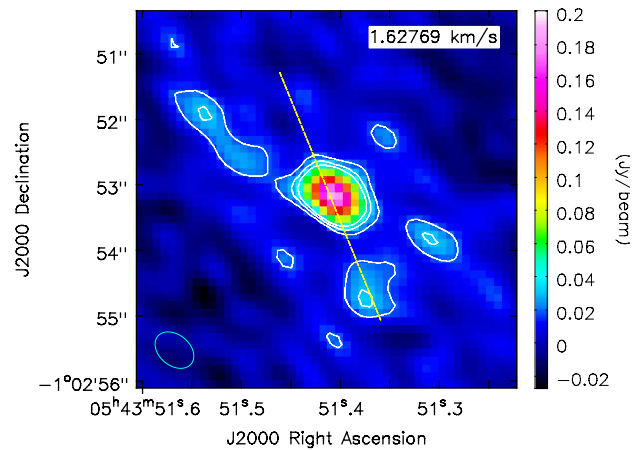}
 \caption{
 HDCO colour map and contour map overplotted for the 1.63 km s$^{-1}$ channel which is close to the systematic velocity
 1.7 km s$^{-1}$. The contours are from 3$\sigma$ ($\sigma \sim$7 mJy) with a step of 2$\sigma$.
    The HDCO emission map at systematic velocity is quite similar to 'X' shaped outflow.}

    \label{fig:hdco_sysve}
\end{figure}

\begin{figure*}
 \includegraphics[width=8cm, angle=270]{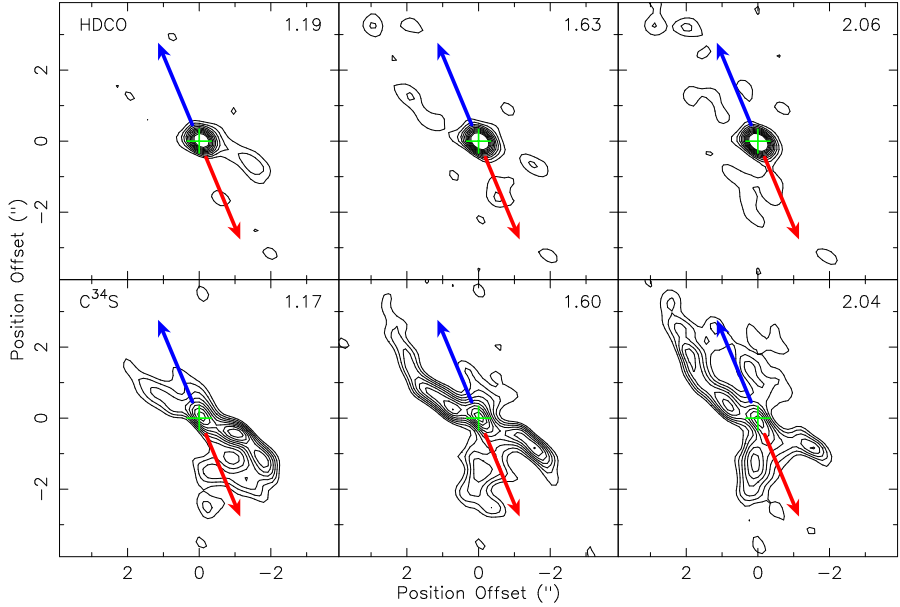}
 \caption{Comparison of channel map of C$^{34}$S and HDCO. Value of velocities are near the systematic velocity 
 $\sim$ 1.7 Km s$^{-1}$. First contour is at 3$\sigma$ and steps are 2$\sigma$ for 
 both HDCO and C$^{34}$S. $\sigma$ is $\sim$ 6mJy/beam and 3.5 mJy/beam for HDCO and C$^{34}$S
 contours, respectively.}
 \label{fig:channel_map}
\end{figure*}

\begin{figure*}
	\includegraphics[width=18cm]{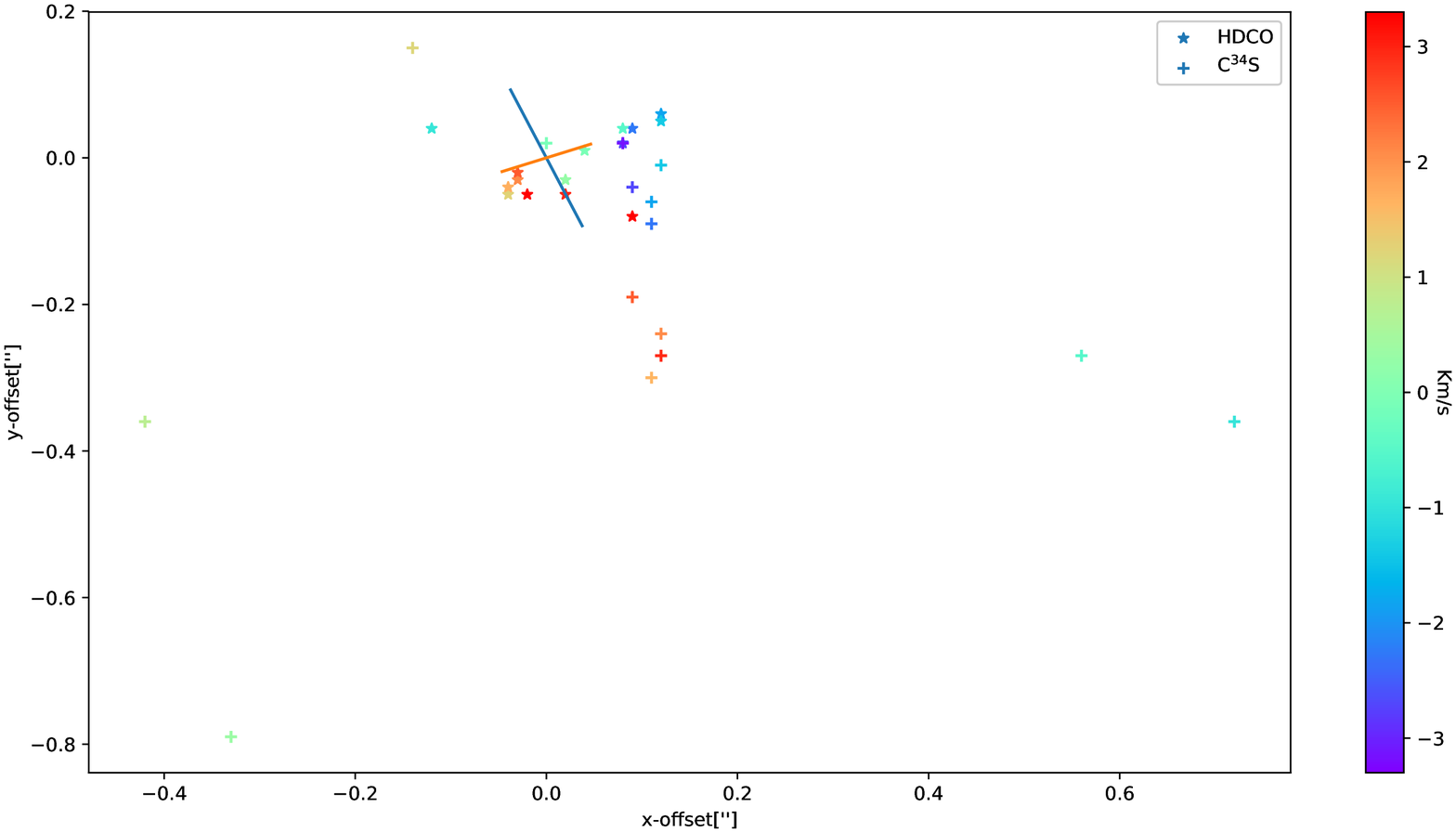}
	\vskip -0.1cm
    \caption{Distribution of the centroid positions of various velocity channels of 
 C$^{34}$S and HDCO line. Velocities are colour-coded according to the 
 color-bar shown in
the figure and value of velocities are subtracted from systematic velocity (1.7 Km s$^{-1}$).
The direction of jet (PA 22) and disk are (PA 112) are shown by lines.  }
    \label{fig:uv1}
\end{figure*}

\begin{figure}
	\includegraphics[width=\columnwidth]{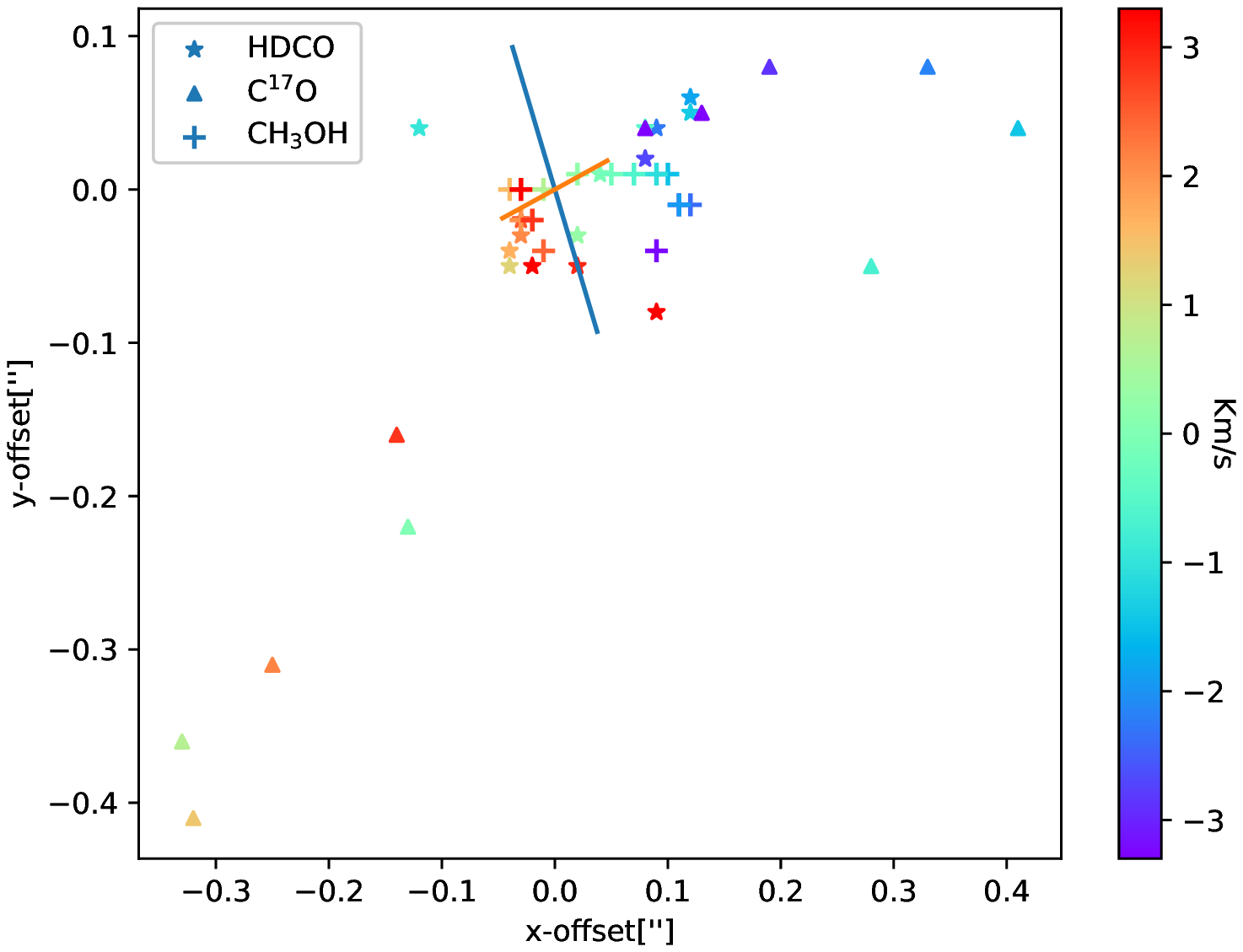}
    \caption{
    Distribution of the centroid positions of various velocity channels of 
 C$^{17}$O, CH$_3$OH and HDCO line. Velocities are colour-coded according to the 
 color-bar shown in
the Figure and value of velocities are subtracted from systematic velocity (1.7 Km s$^{-1}$).
The direction of jet (PA 22) and disk are (PA 112) are shown by lines.
   }
    \label{fig:uv2}
\end{figure}

\subsection{CH$_3$OH, and C$^{17}$O emission}
Methyl alcohol emission of v0 7(1,7)- 6(1,6) transition
is also observed around the MM1 protostar position.
From the spectral signature (Figure~\ref{fig:spectra}) it is seen that CH$_3$OH emission is from
a smaller region than HDCO around the protostar position.
The CH$_3$OH line was identified using CDMS/JPL molecular database, having E$_u$=79K. The kinematics of methanol 
emission is already described by \citet{leurini2016}. 

Methanol emission is from a smaller region  ($0.64'' \times 0.45'' PA \quad 49.6^{\circ}$) 
than that from HDCO emission region. 
In Figure~\ref{fig:uv2} we compare HDCO emission centroid for different velocity channel
with C$^{17}$O and CH$_3$OH. C$^{17}$O
emission traces disk rotation at high velocity and rotating outflow cavity at low-velocity \citep{code2014}. Methanol 
emission shows rotation as the same sense of C$^{17}$O and traces small-scale outflow.
Though the spatial resolution ($0.6 ''$) is not enough to ascertain the fact,
we have compared HDCO emission with these emissions. At small velocities, HDCO emission shows 
some rotation near the disk plane. This rotation may trace small scale outflow or 
a rotating environment \citep{lee2017a} near the disk plane. At high velocities, the red-shifted centroid shifts towards
jet axis and away from disk plane in southern portion of the outflow region. This is very different than the 
methanol velocity centroids at high velocities.  Hence at small velocity it may trace some rotation near
the disk which may be because of small-scale outflow or some rotation environment \citep{lee2017a}.
At systematic velocities HDCO emission traces `X' shaped outflow cavity similar to the C$^{34}$S emission; it 
may trace small scale outflow too but due to limited spatial resolution we are not sure about this.
 
\subsection{Column density calculation and abundances}
We calculate the column densities of C$^{17}$O and HDCO, assuming the lines are optically thin.
As we do not know the rotational temperature of HDCO transition, we consider a range 
of temperature (20 K-300 K) as the excitation temperature of the molecular transition.
Now, 
\begin{equation}
N_u/g_u=\frac{N_{tot}}{Q_{T_{rot}}} e^{\frac{-E_u}{T_{rot}}}
           =\frac{8\pi\nu^2 k\int T_b dv}{hc^3Ag_u}
           =\frac{3k\int T_b dv}{8\pi^3\mu^2\nu S} 
\end{equation}
where, g$_u$ is the statistical weight of the upper level u, N$_{tot}$ is the total 
column density of the molecule,$ Q_{T_{rot}}$is the rotational partition function, $T_{rot}$ is rotational
temperature, $E_u$ is the upper energy level, $k$ is the Boltzman constant, $\nu$ is the frequency 
of the line transition, $A$ is the Einstein co-efficient of the transition. $\int T dv$ is the 
integrated line intensity. Calculated column density is a beam average column density, so we consider the
beam dilution factor, given by,
\begin {equation}
\eta_{BD}=\frac{\theta_s^2}{\theta_s^2+\theta_B^2}
\end {equation}
where, $\theta_S$ is the source size and $\theta_B$ is the beam size. 
As the source is marginally resolved in HDCO emission, we are unsure about the source size.
\citet{code2016} considered  source size $\sim 0.3''$, which is about the size of dusty disk \citep{lee2017b}.
\citet{leurini2016} considered source size  $\sim 0.2''$. Here, we consider source size $\sim 0.2''$.
In that case, the calculated column density would be multiplied by $\simeq 10$ for central 
region only (region `M' in Fig. 6), if we consider beam dilution.
The spectral profile for different region is shown in Figure~\ref{fig:spectra}.
Calculated average column densities without beam dilution correction 
over different regions (Figure~\ref{fig:region}) are enlisted in Table~\ref{tab:colden}.
Consequences of beam dilution effect are discussed section 3.5.
\begin{table*}
	\centering
	\caption{Column density of HDCO around MM1 protostar for different regions depicted in Fig~\ref{fig:region}}
        \label{tab:colden}
	\begin{tabular}{|l| c| c| c| c |c|c|} 
		\hline
		Regions & \multicolumn{5}{c|}{Column densities for various T$_{rot}$ in unit of $10^{14}$ (cm$^{-2}$)}&Error \\
		        &  300 K   & 160 K &  90 K  & 40 K    & 20 K & \\
		\hline        
		M       & 12.3& 5.6 & 3.1  &   2.0  &  ..          &22.8\%\\
		L2      & ..   & 1.0  & 0.5   & 0.36    & 0.52     & 25.6\%\\
		L3      &  ..  & 1.3 & 0.74  &   0.48  & 0.69     &23.6\%   \\
		D2      &  . . & 0.86 & 0.48  & 0.31    & 0.44      &25.2\%  \\
		D3      &  ..  & 1.3  &  0.74 & 0.48    & 0.69     & 25.4\% \\
	\hline
	\end{tabular}
\end{table*}
In the Table ~\ref{tab:colden} we have shown the column 
density for different regions along the jet and perpendicular to the jet axis. 
The central hot-corino is expected to be hot, so we have excluded the lowest temperature (20 K)
in our range of temperature to calculate column density in the central region. Similarly, in regions away from the central
source we have not considered the 300 K temperature for the calculation. Considering the H$_2$ column density 
$\sim 10^{24}$cm$^{-2}$ (see next section), X$_{HDCO} \sim 10^{-10}$. 
The errors for column density 
are calculated for noise and statistical  (Gaussian) fitting. If we consider 
 a calibration error of 20\% then the same uncertainty will be added to column density calculation in 
 addition to the statistical error; the resultant error shown in Table~\ref{tab:colden}.
Due to low signal to noise ratio we have not estimated the column densities in outermost 
regions e.g. L1, D1 in Figure~\ref{fig:region}

\subsection{C$^{17}$O emission and disk mass}

In earlier Section it was mentioned that C$^{17}$O traces a Keplerian disk at high velocities. Assuming
C$^{17}$O emission to be optically thin, we can calculate disk mass from the beam-averaged C$^{17}$O
column density and converting it to H$_2$ volume density. For conversion of C$^{17}$O column density to H$_2$,
we use $X_{CO}/X_{C^{17}O}$=1792  \citep{wilson1994} and $X_{CO}=N_{CO}/N_{H_2}
 \sim 10^{-4}$.
 In region `M' (Figure~\ref{fig:region} ) C$^{17}$O column density for T$_{rot}$ = 90-300 K  is 
 1.3--3.3$\times 10^{16}$ cm$^{-2}$, so N$_{H_2}\sim  2.3-5.9\times 10^{23}$ cm$^{-2}$, as a beam 
 averaged column density. As the disk is not resolved and has been seen edge on, considering beam 
 dilution H$_2$ volume density becomes ${\rm 8\times N_{H_2}/D_{disk}} $ where the disk size is $\sim 90$AU 
 \citep{code2014}. Using the above information H$_2$ volume density becomes 1.3--3.4$\times 10^{9}$cm$^{-3}$.
 Considering H$_2$ volume density $\sim 10^{9}$cm$^{-3}$, we derive the disk mass to be 0.016 M$_\odot$.
 It is close to the value 0.014  M$_\odot$, estimated by \citet{lee2014}.
 Here, we have considered the formula, $M_D \sim 1.4 \times m_{H_2} \times \pi r^2 \times 2H$, where `r' 
is the disk radius and `H' is the disk height from mid-plane; here the factor, 1.4 accounts for 
 the mass in the form of helium. We consider H$\sim 40$AU and the disk radius to be 90 AU \citep{lee2014}. The  
 disk mass can also be calculated from continuum emission. We have considered that the continuum from HH212 disk is
 optically thin and isothermal. Though the disk mass is not constant as suggested by \citet{lee2014}, we have 
 calculated it by assuming a dust temperature to be $\sim$ 90 K.
 Following  earlier work \citep{tobin2012}, we assume the spectral index
 $\beta$=1 \citep{kwon2009}, and the dust opacity $\kappa_0$=0.035 g cm$^{-2}$ at 850 $\mu$m \citep{andrew2005}.
 We use the following formula \citep{tobin2012} to calculate the disk mass
 $$ M_{dust}=\frac{D^2F_\lambda}{\kappa_0 (\frac{\lambda}{850\mu m})^
 {-\beta}B_\lambda(T_{dust}).}$$
 With {T$_{dust}$=90 K}, the disk mass is 0.0159 M$_\odot$, which is close to 0.016 M$_\odot$, 
 the disk mass calculated based on hydrogen mass.

\subsection{Discussion}
\subsubsection{chemistry}
Formaldehyde can be formed in both the gas and the grain phase. On grain surfaces, 
formaldehyde forms through sequential reactions of H or D atom with CO:\\
CO $\rightarrow$ HCO $\rightarrow$ H$_2$CO and 
CO $\rightarrow$ DCO $\rightarrow$ HDCO/D$_2$CO  \citep{cazaux2011, taquet2012}. The grain phase reactions
occur mainly in cold temperature condition ($<$50K). Gas phase formation of formaldehyde and its deuterated form is
also possible through reaction involving CH$_2$D$^+$. More specifically, at relatively high temperatures, T$\sim$ 
100K or higher, this reaction is relevant to the central hot-corino region of HH212
\citep{wootten87, oberg2012}. Reaction involving CH$_2$D$^+$ is not active in cold region due to 
its high exothermicity ($\Delta$E of 654 K, see \citealt{roueff13}) but in high temperature region this reaction 
may take part and favours deuterium fractionation in gas-phase:

CH$_3^+$ + HD$\rightarrow$ CH$_2$D$^+$ + H$_2$ + $\Delta$E.

\citet{fontani14} first time disentangled the emission of deuterated formaldehyde 
form on grain surfaces from its gas phase production. In our observation 
we have not seen any H$_2$CO line but still we can infer about the production region and possible production
route of deuterated formaldehyde (HDCO). HDCO emission is from two regions, one is from central hot-corino
region another is from outflow cavities. The impact of bipolar jet on the cavities may release HDCO from grain
surfaces due to sputtering or sublimation from grain mantle (e.g., see \citealt{code2012}). As the regions in 
the outflow cavities have low-temperature, hence we can expect the production of HDCO is mainly from grain 
surfaces. If we consider the column density of formaldehyde in central region assuming a temperature 160 K and 
40 K in one of the outflow region (south-west lobe, centered at $\alpha(J2000)= 05^h43^m51^s.37.4,
 \delta(J2000)=-01^{\circ}02'54''.69$; see Fig~\ref{fig:hdco_sysve}),
then the column densities are respectively 5.6$\times 10^{14}$
and 0.18$\times 10^{14}$ cm$^{-2}$. In the outflow region this is a factor $\sim 31$ less than the central 
region column density of HDCO.
From C$^{17}$O emission we can get a rough idea about the density difference in these two regions. In the same 
region (south-west lobe) we have calculated the C$^{17}$O column density, and it is a factor of
10 less than the centre region (`M'). Hence, if HDCO comes from only grain surfaces then their column densities should be 
decreased similarly by the density factor. Considering the density differences, the column density in the 
central region is a factor of three ($\sim 31/10$) higher.
Hence, indirectly it implies that there is active gas phase HDCO production in central hot-corino region
along side the grain-phase desorbed HDCO. Recently, another doubly deuterated formaldehyde (D$_2$CO) has been 
reported by \citet{lee2017a} for the same source for a beam size of $\sim 0.04''$. In this work, HDCO line emission 
is marginally resolved, if we consider emission region $0.2''$ then considering the beam dilution,  a factor of 10 
 will be  multiplied by central region column density. In that case, 
the column density of HDCO and D$_2$CO would be comparable ($\sim 10^{15}$). 
If we consider smaller regions of the emission region such as the D$_2$CO beam,  then the column density 
will be higher for HDCO. Hence, considering the source size $0.2''-0.04''$, the D$_2$CO/HDCO ratio becomes
$1-0.04$. We guess the emission region for HDCO will not be as small as that for D$_2$CO, in that case the deuteration of formaldehyde 
in quite higher than methanol for the same source. Methanol deuteration (D/H) for the HH212 is 2.4$\pm 0.4 \times 10^{-2}$
as reported by \citet{bian2017}. On the other hand  \citet{lee2017a} suggested D/H
$\sim 0.27$ for methanol in the disk environment. If we consider D/H ratio 0.27 to be true
then methanol is still produced effectively on the grain surface in the disk environment. \citet{bian2017} used
 $\rm{^{13}CH_3OH}$ and LTE approximation to calculate methanol column density. If we consider D/H ratio
of 2.4$\pm 0.4 \times 10^{-2}$ to be more likely then probably due to high temperature deuterated 
methanol production is not efficient in the hot-corino region. We speculate that deuterated formaldehyde can still be
produced in this region through gas phase reaction network unlike the methanol production. 
\subsubsection{kinematics/morphology}
\begin{figure}
	\includegraphics[width=\columnwidth]{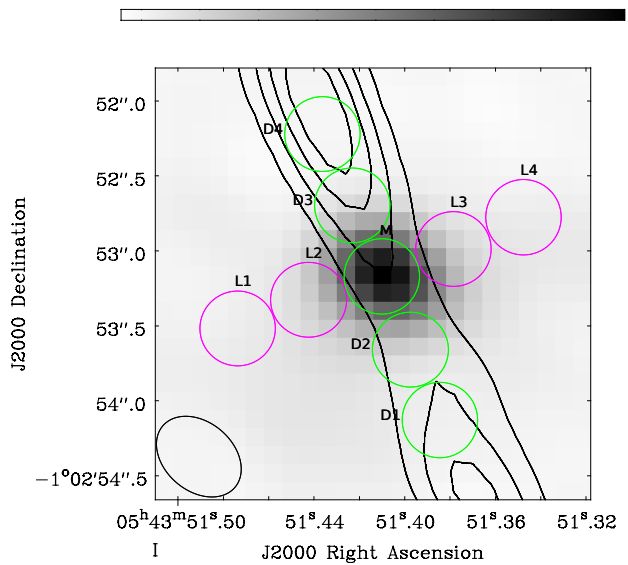}
    \caption{Different region along and perpendicular to the jet axis. The region circle's diameter is 0.5 arcsec
    similar to the beam width. The gray scale image is the continuum image of the central source, namely, HH212. 
The contours are for SiO jet emission along a PA 22$^o$.}
\label{fig:region}
    \end{figure}

\begin{figure*}
	\includegraphics[width=12cm]{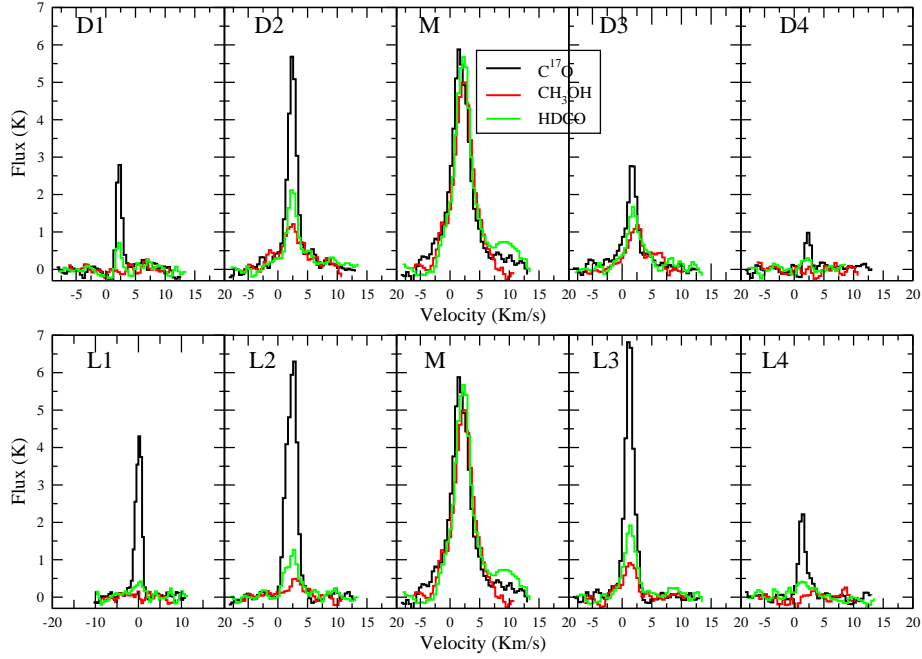}
    \caption{Spectral profile in different regions as shown in Figure~\ref{fig:region}.}
    \label{fig:spectra}
\end{figure*}

 In Figure~\ref{fig:region} we have 
defined different circular regions ($0.5''$ diameter) comparable to the beam size and the observed spectra
towards these regions are shown in Figure~\ref{fig:spectra}. We can see from 
Figure~\ref{fig:uv1},~\ref{fig:uv2} there may be emission from rotating environment or small-scale outflow near the disc 
but it is very uncertain. At high velocity the emission is certainly shifted above the disk along the jet
which signifies that the emission is affected by outflow or disk-wind at the base of the jet.
The column density (Table~\ref{tab:colden}) 
on both sides (L2, L3) of the central circular region of $0.5''$ diameter is less than that at the central region. 
We can see a jump in column density for HDCO. This is due to the fact that 
at the disk ($\sim 90$AU) and envelope interface there is a sharp rise of temperature
and density due to accretion shock. Also, from Figure~\ref{fig:spectra}
we see that the spectral signature of methanol emission is absent from the outermost regions (L1, L4, D1, D4)
though a weak signature of HDCO emission is present. From this also we can say that HDCO emission is more extended than 
methanol. \citet{lee2017a} at a scale of $0.04''$ found that difference of emission regions of the complex
molecules is partly due to different A-coefficients. In that observation of HH212, it was found that molecules
with comparatively low A-coefficients are seen to be more extended than the that with high A-
coefficients. Here, A-coefficient of methanol transition is lower than that of HDCO, but the region seems 
to be more compact than that of HDCO. Higher A-value may correspond to higher critical 
density. \citet{guzman2011} described H$_2$CO critical densities $\sim 10^6$; 
due to high density ($\sim 10^8-10^9$ cm$^{-3}$) in the central region it is expected that LTE condition 
is maintained for the reported molecular transitions here.
Hence, we can speculate that it is due to local physico-chemical condition and high line 
strength, comparatively lower E$_u$ of HDCO which may be responsible for
the emission difference. Another difference we can see for methanol and HDCO emission in central region (`M') is this:
HDCO has a red-shifted peak which is absent in CH$_3$OH emission. This peak may be from the line contamination of other 
molecular transition, but we have not found any such line from other molecules. Alternatively, this may be 
related to the high velocity outflow from the base of the jet. 

\section{Conclusions}
In this letter we described the emission of deuterated formaldehyde (HDCO) from the hot inner region 
of HH212. This emission is limited mainly to the inner $\sim 200$ AU. The kinematics of 
HDCO is quite similar to that of the C$^{34}$S emission \citep{leurini2016} near systematic velocity.
HDCO traces large scale outflow cavity near systematic velocity; 
it may trace small-scale outflow or disk wind
but due to limitation of spatial resolution in this observation we are uncertain about this.
On the other hand, both methanol and HDCO emission peak values are $\sim 1.9$ km s$^{-1}$ 
and the spectral profile is symmetric in low to medium velocity range ($<2.4 {\rm km s^{-1}}$). 
The asymmetry at high velocity for HDCO may be associated with the outflow near the disk plane.
The HDCO rotation may be associated with the disk wind or rotating environment \citep{lee2017a};
or to the rotating cavity wall, similar to C$^{34}$S. Due to limited resolution of the observation
we cannot conclude about the rotation with certainly. The emission 
is assumed to be optically thin. Here we observe only one transition of HDCO.  Thus we cannot 
determine the excitation temperature. We have considered a range of possible temperature depending on earlier 
studies and typical hot-corino temperature assumed in the literature. 
The column density of HDCO is $\sim 10^{14} {\rm cm^{-2}}$. Though 
we have not observed any H$_2$CO transition but comparing results of D$_2$CO given in \citet{lee2017a} 
we speculate that the deuterium fractionation of formaldehyde is relatively higher than methanol in the central region.
We guess that the gas phase formation of deuterated formaldehyde is active in the central hot region in the low-mass 
protostar HH212.
\section*{Acknowledgements}
We acknowledge the anonymous referee for the constructive comments.
We also thank Dr. Nirupam Roy (IISc, India) for his helpful suggestions.
DS is thankful to Department of Space, Govt. of India (PRL) for support in continuing research and  also 
want to thank Young Visitor Programme (KASI) for financial help to work in Korea Astronomy
Space science Institute for a short time. AD want to acknowledge ISRO Respond (Grant no. ISRO/RES/2/402/16-17).
C. -F.L. acknowledges grants from the Ministry of Science and Technology of Taiwan (MoST 104-2119-M
-001-015-MY3) and Academia Sinica (Career Development Award).
This paper makes use of the following ALMA data: ADS/JAO.ALMA\#2011.0.00647.S.
ALMA is a partnership of ESO (representing its member states), NSF (USA) and NINS (Japan),
together with NRC (Canada), NSC and ASIAA (Taiwan), and KASI (Republic of Korea),
in cooperation with the Republic of Chile. The Joint ALMA Observatory is operated by ESO, AUI/NRAO and NAOJ.











\bsp	
\label{lastpage}
\end{document}